# TOWARDS INTERNET OF THINGS (IoTs): INTEGRATION OF WIRELESS SENSOR NETWORK TO CLOUD SERVICES FOR DATA COLLECTION AND SHARING

Rajeev Piyare[1] and Seong Ro Lee[2]

[1,2] Department of Information Electronics Engineering, Mokpo National University, 534-729, South Korea
`rajeev.piyare@hotmail.com;srlee@mokpo.ac.kr`

## ABSTRACT

*Cloud computing provides great benefits for applications hosted on the Web that also have special computational and storage requirements. This paper proposes an extensible and flexible architecture for integrating Wireless Sensor Networks with the Cloud. We have used REST based Web services as an interoperable application layer that can be directly integrated into other application domains for remote monitoring such as e-health care services, smart homes, or even vehicular area networks (VAN). For proof of concept, we have implemented a REST based Web services on an IP based low power WSN test bed, which enables data access from anywhere. The alert feature has also been implemented to notify users via email or tweets for monitoring data when they exceed values and events of interest.*

## 1. INTRODUCTION

The Internet of Things (IoTs) can be described as connecting everyday objects like smart-phones, Internet TVs, sensors and actuators to the World Wide Web where the devices are intelligently linked together enabling new forms of communication between things and people, and between things themselves. Building IoTs has advanced significantly in the last couple of years since it has added a new dimension to the world of information and communication technologies. According to [1], in 2008, the number of connected devices surpassed connected people and it has been estimated by Cisco that by 2020 there will be 50 billion connected devices which is seven times the world population. Now anyone, from anytime and anywhere can have connectivity for anything and it is expected that these connections will extend and create an entirely advanced dynamic network of IoTs. The development of the Internet of Things will revolutionize a number of sectors, from wireless sensors to nanotechnology.

In fact, one of the most important elements in the Internet of Things paradigm is wireless sensor networks (WSNs). WSNs consist of smart sensing nodes with embedded CPUs, low power radios and sensors which are used to monitor environmental conditions such as temperature, pressure, humidity, vibration and energy consumption [2]. In short, the purpose of the WSN is to provide sensing services to the users. Since, the number of users of the Internet is increasing therefore; it is wise to provide WSN services to this ever growing community.

Cloud computing is a flexible, powerful and cost-effective framework in providing real-time data to users at any time with vast coverage and quality. The Cloud consists of hardware, networks, services, storage, and interfaces that enable the delivery of computing as a service [3]. In addition, it's also possible to upload the data obtained from the wireless sensor nodes to the Web services based on Simple Object Access Protocol (SOAP) and Representational State Transfer (REST), using messaging mechanisms such as emails and SMS or social networks and blogs [4]. By connecting, evaluating and linking these sensor networks, data conclusions can be made in real-time, trends can be predicted and hazardous situations can be avoided.

In this paper, we present the design, development and integration of an extensible architecture for WSN with the Cloud based sensor data platform, Open.Sen.se [5] where info-graphic of different data streams can be displayed, accessed and shared from anywhere with Internet connectivity. The collected data from the sensor nodes are processed, stored and analyzed on Open.Sen.se server via an Application Programming Interface (API). We have used REST based Web services as an interoperable application layer that can be directly integrated into other application domains like e-health care services, smart homes, or even vehicular area networks (VAN). For proof of concept in a smart environment, we have implemented a REST based Web services on an IP based low power WSN test bed, which enables data access from anywhere for the smart environment.

The remaining of the paper is organized as follows. In Section 2, we briefly discuss related work. Section 3 describes the proposed architecture while Section 4 outlines the hardware design of the base station and the End Nodes. In Section 5 we discuss the software implementation of our approach. Section 6 presents the implementation results and discussions and finally, some conclusions are presented.

## 2. RELATED WORK

Wireless sensor platforms have been widely deployed in a number of applications ranging from medical such as Alarm-Net [6], or CodeBlue [7] to environmental monitoring [8-10]. The architecture of these systems has been designed in a very ad hoc fashion and is not flexible to adapt to other applications or scenarios while the core problem is the same, remote monitoring using sensor networks. During the last few years, many researchers have investigated on ways to connect wireless sensor networks to the Cloud [11]. Authors in [12-16] have presented Internet protocols for connecting wireless sensor networks to the Internet but no real implementations have been shown. Much of the previous work has been on theoretical aspects of system architecture rather than actual deployment and testing of wireless sensor networks with the Clouds. Use of Web services to connect sensor networks with external networks have also been suggested by researchers in [17, 18]. However, their work was mainly focused on the feasibility of SOAP based Web services in terms of energy and bandwidth overheads.

SenseWeb [19] is one of the first architectures being presented on integrating WSN to the Internet for sharing sensor data. Users were able to register and publish their own sensor data using the SenseWeb API. The main drawback of SenseWeb is that all the decision making process is executed at a single central point called the Coordinator. The Coordinator is the central point of access for all applications and sensor contributors where all the sensor data is stored and analyzed. That is, all the intelligence to control and to make a decision is located at this central point and if the Coordinator fails, the entire network is disrupted.

It is therefore suggested that the various decision levels can be implemented onto different architectural layers. The upper level known as Supervision Layer will be used for all sensor data

International Journal of Computer Networks & Communications (IJCNC) Vol.5, No.5, September 2013

storage, analysis and for decision making, while the Sensor Layer where the sensors are deployed can be used to partially analyze data and the determination of reactive response. The Coordinator still exists as a point of control for analysis of data and remote monitoring as well as acting as a gateway between sensors and the Cloud.

In order to address the above mentioned issues of flexibility and centralized decision making process, we designed and implemented a more flexible architecture for integrating WSN to Cloud using REST based Web services as an interoperable application layer which can be directly integrated into other applications. The architecture presented in this work can be customized in different ways in order to accommodate different application scenarios with minimum recoding and redesign. To build a low power and self-healing Wireless Sensor Network we have used XBee ZB modules which are ZigBee-complaint wireless sensor networking devices developed by Digi International, Inc [20]. Due to low power, simple network deployment, reliable data transmission and low installation costs, the ZigBee wireless standard has been preferred for this study over Wi-Fi and Bluetooth. In addition, to reduce the overall cost of implementation and network latency, each End Device is only equipped with an XBee ZB module with sensors. Furthermore, to reduce energy consumption and to increase the network lifetime, sleep mechanism for battery powered sensor nodes have been utilized.

## 3. DESCRIPTION OF PROPOSED ARCHITECTURE

The architecture of the proposed system is divided into three layers (Figure 1): Sensor Layer, the Coordinator Layer and the Supervision Layer.

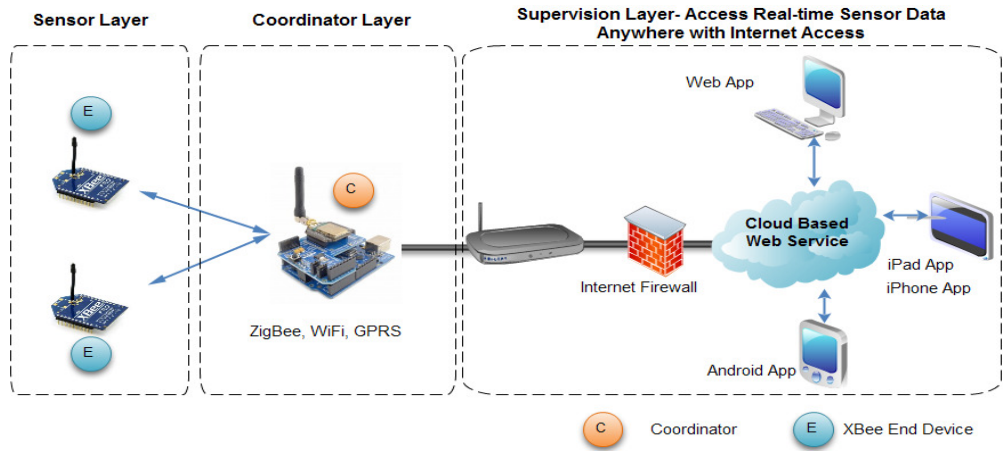

Figure 1. Proposed Architecture

The Sensor Layer consists of sensors that interact with the environment. Every sensor was integrated with wireless nodes using an XBee ZB platform called End Devices. These End Devices form a Mesh network and send the information gathered by the sensors to the Coordinator Layer through the sink node called the base station. Messages are routed from one End Device to another until they reach this base station. There are several hardware platforms available for wireless sensor network deployment such as TelosB, Mica, IRIS and Wasp mote. For our prototype system, we have utilized XBee module from Digi International, Inc. Each XBee ZB module has the capability to directly gather sensor data and transmit it without the use of an external microcontroller, a capability known as XBee direct [21]. This offers many advantages. By using XBee alone, it can minimize weight which is an important factor for systems such as Body Sensor Networks or wearable's. Omitting an external microcontroller also reduces power consumption which is a critical advantage for wireless systems that run on batteries and save money. However, there are also some important tradeoffs associated with this.

The Coordination Layer is responsible for the management of the data received from the sensor network. It temporarily stores the gathered data into buffer and sends it to the Supervision layer at predefined intervals. Base station which comprises of Arduino UNO, Ethernet shield and XBee is connected to the Internet using RJ45 cable and is powered using an AC adaptor. It serves as a mobile mini application server between the wireless sensors and the dedicated network and has more advanced computational resources compared to the End Devices found in Sensor Layer. At the base station, the sink node gathers data from wireless sensors using the ZigBee protocol and sends this data to Cloud based sensor data platforms.

Finally, the Supervision Layer accommodates the base station with a Web server to connect and publish the sensor data on the Internet. This layer stores the sensor data in a database and also offers a Web interface for the end users to manage the sensor data and generate statistics. For the Supervision Layer, we have used Open.sen.se [5] HTTP Service which provides a REST based API to publish and access the sensor data. Thus, allowing existing networks to be connected into other applications with minimal changes. Open.Sen.se offers a graphical interface for real-time monitoring of systems using info graphic data streams and to retrieve the sensor values using device type and timestamp. Alerts can also be automatically generated to notify the user each time if the desired event has been sensed by the domain rules programmed in the base station.

## 4. WIRELESS SENSOR NODE DESIGN

This section highlights the design and development phases of test bed in terms of hardware in order to integrate it to the proposed architecture.

### 4.1. Base station (Coordinator) Design

The base station plays a key role in our proposed system as illustrated in Figure 1. This node has been kept minimum size while ensuring all functions of communication, sensing and calculation. The prototype of the base station is shown in Figure 2. The hardware of base station consists of an Arduino UNO board, an Ethernet shield and an XBee shield that supports XBee ZB module. The Arduino is an open-source microcontroller that uses ATMEGA 328, an Atmel AVR processor which can be programmed by the computer in C language via USB port [22]. Arduino also has on-board 5 analog pins and 13 digital pins for input and output operations, supporting SPI and I2C which can be used to interface with other devices. The role of the microcontroller in this wireless sensor network is to collect sensor readings from the End Devices via XBee ZB module, arrange sensor data using developed packet protocols and send it to Open.Sen.se server via an Ethernet module. The Ethernet module acts as a central node to bridge the wireless sensor network with local proxy. Generally, the function of the base station is divided into two parts: Web-Server and XBee interface to the wireless sensor network. These two functions are implemented on Arduino UNO. The Web-Server function uses <Ethernet.h> library, while XBee interface uses <XBee.h> library.

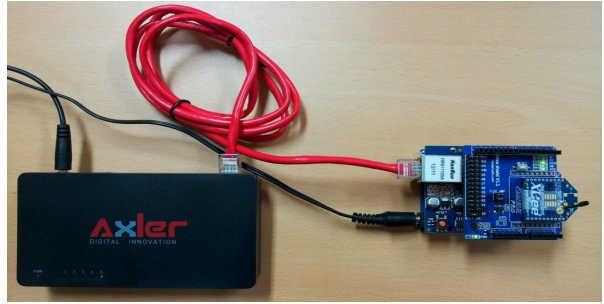

Figure 2. Base station connected to Ethernet router consisting of an Arduino UNO, an Ethernet Shield and XBee ZB module

### 4.2. End Device

#### 4.2.1. XBee ZB Interface

The End Devices consist of two parts: Sensor Interface and XBee ZB Interface as illustrated in Figure 3. The End Device is developed based on XBee Radio Frequency module operating in an unlicensed band of 2.4 GHz with a data transfer rate of 250 kbps [23]. XBee uses ZigBee protocol and support the needs of low cost, low power wireless sensor networks. ZigBee is built on top of IEEE 802.15.4 standard which defines the Medium Access Control (MAC) and physical layers. ZigBee protocol also features multi-hop communication capability, therefore providing a vast range of communication and a wide coverage area [24]. An XBee ZB offers transmission range of 40m in indoor scenarios and 140m in outdoor. End Devices wait for data reading request (i.e. Polling) from the Coordinator and then responses with the value from the sensor. Polling is a method in which the network Coordinator requests each End Device one by one to send sensor readings. This avoids interference from multiple nodes transmitting to the Coordinator simultaneously.

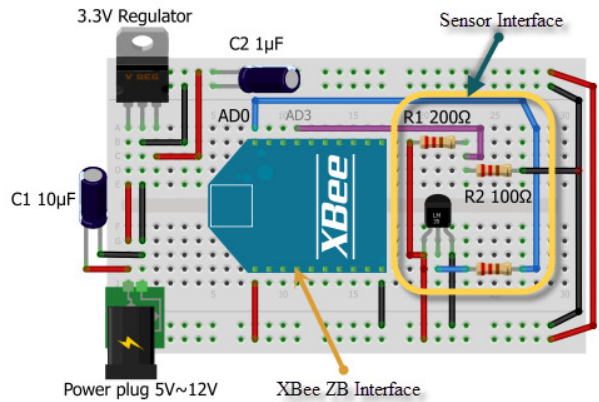

Figure 3. End device with temperature and voltage sensing unit

#### 4.2.2. Sensor Interface

A cost reduction for each node is achieved by removing the additional use of a microcontroller and using XBee ZB as a standalone device known as XBee direct as mentioned in Section 3. Since, XBee houses on-board 9 analog and digital input and output pins, sensors can be directly interfaced to it. This allows XBee modules to automatically sample the sensor inputs and report back to the Coordinator using API firmware. There are three End Devices for this experimental setup and each consists of a temperature sensor. For the temperature monitoring sensor, we have used a low cost LM35 analog sensor from DF Robot to show the proof of concept. LM35 is a precision integrated-circuit temperature sensor, whose output voltage is linearly proportional to Celsius temperature [25]. This sensor employs a 3-wire interface, has a low impedance and power consumption of 60μA from its supply. The sensor interface reads temperature strings from LM35 on analog pin AD0 of XBee ZB module and sends this data packet to the base station. At the base station, Arduino microcontroller receives this data packet, converts it into numerical values with specific data format and End Device ID. Analog samples are returned as 10-bit values from the XBee ZB modules. This analog signal is then sampled and quantized at the base station by the Arduino into a digital value in the range of 0-1023, where 0 represents 0V and 1023 represents 5V. To convert the A/D reading to mV, the following equation is utilized:

$$AD(mV) = (A/D\ reading \times 1200 mV)/1023 \qquad (1)$$

Then the temperature value in volts is further converted into degree Celsius according to equation (2). Since the scale factor for LM35 is 0.01V/°C, therefore:

$$Temp\ in\ °C = (Vout\ in\ mV)/10 \qquad (2)$$

To monitor the supply voltage for each End Node, voltage sensing unit has also been incorporated. If the voltage level is too low, then the End Device enters sleep mode automatically and a notification is automatically generated and send to the user by Open.Sen.se. One of the main challenges in integrating voltage sensor into an XBee module is matching the output of the voltage to the analog input. XBee analog inputs cannot read more than 1.2V [23]. Therefore, a voltage division circuitry was constructed to map the supply voltage to a safe level for an XBee input (Figure 4).

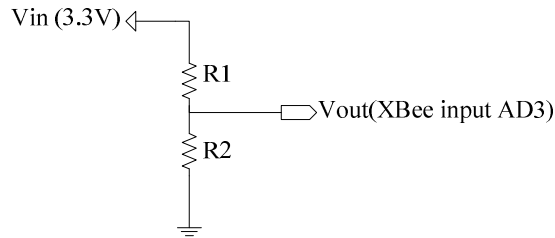

Figure 4. Voltage sensing circuitry

XBee ZB modules can operate within a supply voltage of 2.1V to 3.6V. In order to match the voltage to 1.2V, the values of R1 and R2 were calculated using the voltage division equation:
Where $V_{out}$ is the desired output voltage and $V_{in}$ is the input voltage to the circuit. Using $V_{in}$ as

$$V_{out} = \frac{V_{in} \times R2}{R1 + R2} \qquad (3)$$

3.3V, R1 as 200Ω, R2 was found to be 100Ω. This provides a voltage output of 1.1V, which is within the tolerance of XBee. This voltage data is then transmitted to the base station using the method as described above. At the base station this data packet is again converted into numerical values based on the following equation:

**4.2.3. Power Supply**

$$AD(mV) = ((A/D\ reading \times 1200 \times 3)/1023)/1000 \qquad (4)$$

International Journal of Computer Networks & Communications (IJCNC) Vol.5, No.5, September 2013

The prototype sensor nodes are powered by a 2000 mAh, 9V Energizer lithium polymer battery. This particular battery was preferred because of its long battery life and is rechargeable, which is of interest for continued deployment.

## 5. SOFTWARE DEVELOPMENT

Different software products were developed for this wireless sensor network experiment in order to establish the sensor interface, configure ZigBee network and manage the sensed data for receiving, storing and publishing it on Cloud. Each development phase is described as follows.

### 5.1. XBee Module Configuration

To realize the proposed network architecture, XBee ZB modules were configured to behave as Coordinator and End Devices. XBee supports two modes of operation: Transparent mode (AT) and Application Programming Interface (API) mode with the Escape (ESC) character. API mode was chosen for this research due to following reasons:

1. Allows XBee modules to receive input and output data from one or more remote XBees.
2. MAC layer Acknowledgment (ACK) and retries. This ACK packet indicates to the source node that the data packet was successfully received by the destination node. If ACK is not received, the source node will resend the packet.
3. Receive packets contain the source address of the transmitting node.
4. Packets include a checksum for data integrity.

The data frame for API operation is shown in Figure 5 which is divided into four sections; Start Delimiter, Length section, Frame Data and Checksum. The checksum is calculated as below:

$$Checksum = 0 \times FF - \sum of\ all\ bytes\ in\ API\ structure \qquad (5)$$

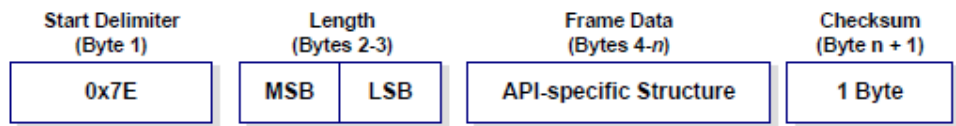

Figure 5. XBee API data frame

### 5.2. Communication and Sensor Layers in Arduino

To successfully communicate with remote sensor nodes from the base station, communication and sensor layers have been implemented on the Arduino. The libraries in the communication layer are used to establish a reliable connection between the sensor nodes and to communicate with Open.Sen.se server. The <XBee.h> libraries are used to receive data on Arduino and create output messages in JavaScript Object Notation (JSON) format. Figure 6 shows the flowchart of communication and sensor layers in Arduino and the End Nodes. A base station is connected to Open.Sen.se server over TCP/IP. Since Arduino Ethernet shield already supports a TCP/IP stack, we have focused on implementing software to connect it to Open.Sen.se server. When Arduino is turned on, it first connects to a local server using a static IP address. To optimize the process of connection, we have used static IP address rather than acquiring an IP via Dynamic Host Configuration Protocol (DHCP). Once the connection is successful, the Coordinator requests for the data from the End Devices. Upon successful reception of data packets, it's decoded and converted into numerical values as described in Section 4.2.2. These values are then updated on Open.Sen.se platform using GET and POST HTTP method which is described in detail in the next section. Since the Open.Sen.se server accepts several TCP connections while communicating, it is scalable for the large number of concurrent users. The base station does not need to re-establish TCP connection every time it sends a message

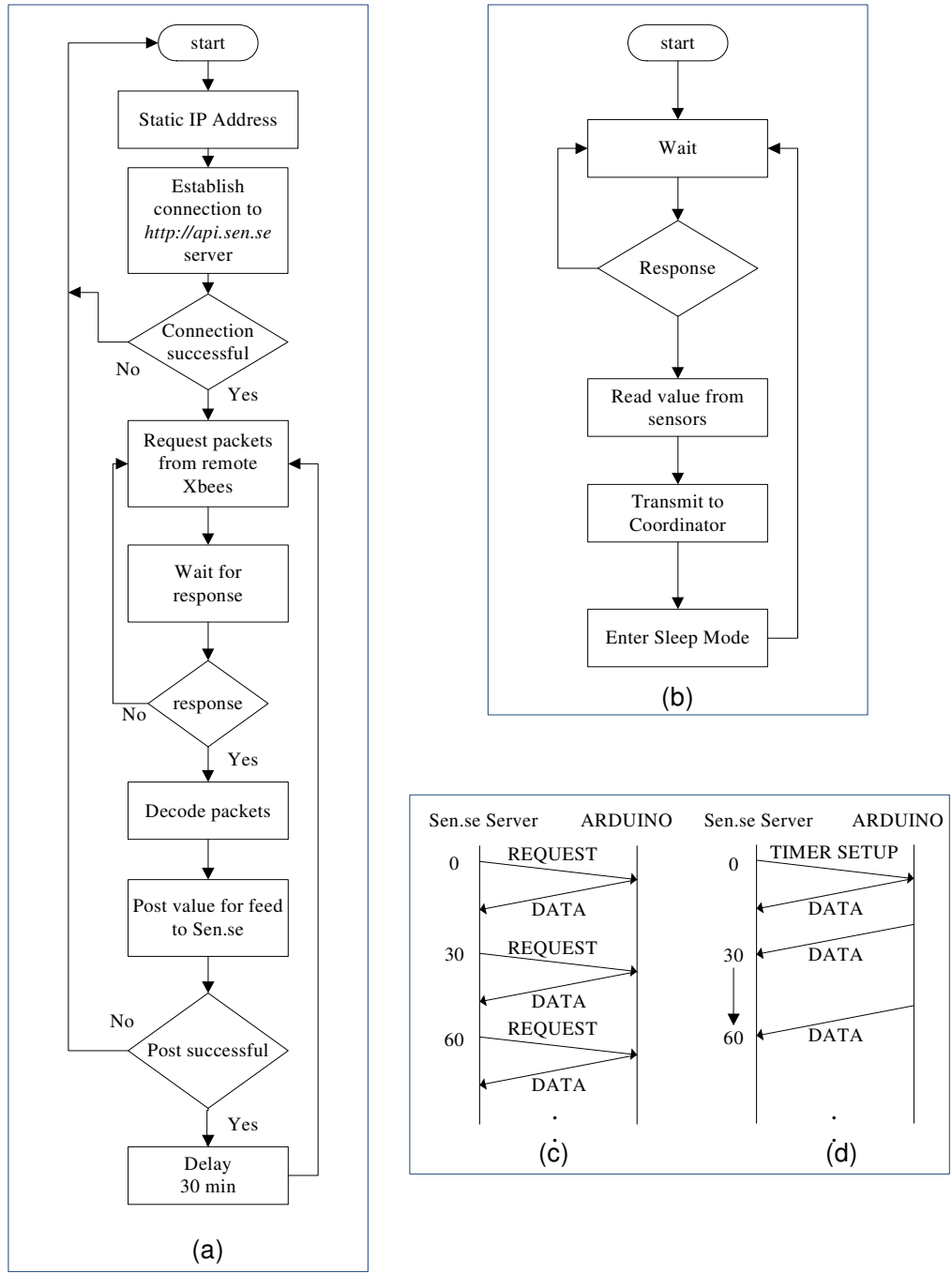

Figure 6.  Flowchart of communication and sensor layers in (a) Arduino and (b) the End Nodes (c) Arduino without Timer and (d) with Timer

### 5.3. Connecting Sensor Network to Cloud Service

As mentioned previously, the access to Cloud services has to be easy, direct, open and interoperable. That is, the provided communication means and programming interfaces (APIs) shall be easy to implement on every platform and developing environment [26]. The most open and interoperable way to provide access to remote services or to enable applications to communicate with each other is to utilize Web services. There are two classes of Web services: Simple Object Access Protocol (SOAP) and Representational State Transfer (REST). REST is a much more lightweight mechanism than SOAP offering functionality similar to SOAP based Web services.

Open.Sen.se is an open source "Internet of Things" application and API to store and retrieve data from things and sensors using Hypertext Transfer Protocol (HTTP) over the internet or via a Local Area Network (LAN). In addition to storing and retrieving numeric and alphanumeric data, Open.Sen.se API allows for numeric data processing such as time scaling, averaging, median and summing. The channel feeds supports JavaScript Object Notation (JSON), Extensible Markup Language (XML), and comma-separated values (CSV) formats for integration into applications.
Therefore, in our approach we have used REST based Web service utilizing standard operation such as GET and POST requests that return (JSON) responses to communicate between the base station and the Open.Sen.se server. JSON is a lightweight data-interchange format. It is easy for human beings to read and write. It is also simpler for machines to parse and generate messages than using XML. For example, to read the current sensor value, an HTTP GET request is sent to the resource of the sensor. The response includes a textual representation of the current sensor value. A soon as the Coordinator decodes the received data packets from the End Devices, an HTTP POST request is sent from the base station to a pre-specified URL, containing the updated value as illustrated in Figure 7. To access the Open.Sen.se API, the following base URL is used: *http://api.sen.se*. Each data entry is stored with a date and time stamp and is assigned a unique Entry ID. In terms of authentication, every communication between the connected Device and Open.Sen.se server is protected with a Sen.se key which is specific and unique to each user.

```
117.17.80.199
117.17.80.1
init done
Received I/O Sample from: 13A2004090C2679
Sample contains analog data
POST value for feed indoor-temperature
Posting in feed indoor-temperature
In postSense
POST in feed indoor-temperature succeeded
POST value for feed node-voltage
Posting in feed node-voltage
In postSense
POST in feed node-voltage succeeded
```

Figure 7. A serial Monitor window showing successful POST for data values

### 5.4. Timer and Reset Function

A timer function is also associated to send notifications to Open.Sen.se server from Arduino periodically. For example, when Sen.se server is required to receive sensor values after every 30 minutes from Arduino, this function is called to configure the Arduino. Once the timer is activated, Arduino reports Open.Sen.se server with the measured sensor data by periods without any further request as illustrated in Figure 6 (c) and (d).

International Journal of Computer Networks & Communications (IJCNC) Vol.5, No.5, September 2013

A reset function initializes all setups on Arduino in software. It performs the same function when the reset button on the Arduino is pressed. If there are conflicts on the communication with Open.Sen.se server, the Arduino will be reset and try to connect with Open.Sen.se server again. Also, Arduino itself calls this function when it finds exceptional errors while connecting.

## 6. IMPLEMENTATION RESULTS AND DISCUSSION

In order to evaluate and demonstrate the proposed model, we implement it by using the technical approach which is described in the above sub-sections. A WSN was created to collect temperature and battery voltage readings. Preliminary experiments were performed to evaluate the system in terms of sensor data accessibility, alert notification time, and battery consumption. Furthermore, Senseboard was created on Open.sen.se server to present the collected data to the user in an easy and meaningful way.

### 6.1. Senseboard

Open.Sen.se server offers graphical interface called 'Senseboard' where different apps can be added. This allows info graphic data streams to be displayed and viewed in real-time anywhere and on any website. It also offers critical multiviz functionality to combine data from multiple sensors into one graph. Figure 8 shows the real-time acquisition curve with measurements showing environment temperature (Red Line) and End Device battery voltage (Green Line). The Senseboard created for this implementation is supported by Internet Explorer, Safari, Firefox, Opera browsers and can also be accessed at http://open.sen.se/sensemeters/tab/3114/.

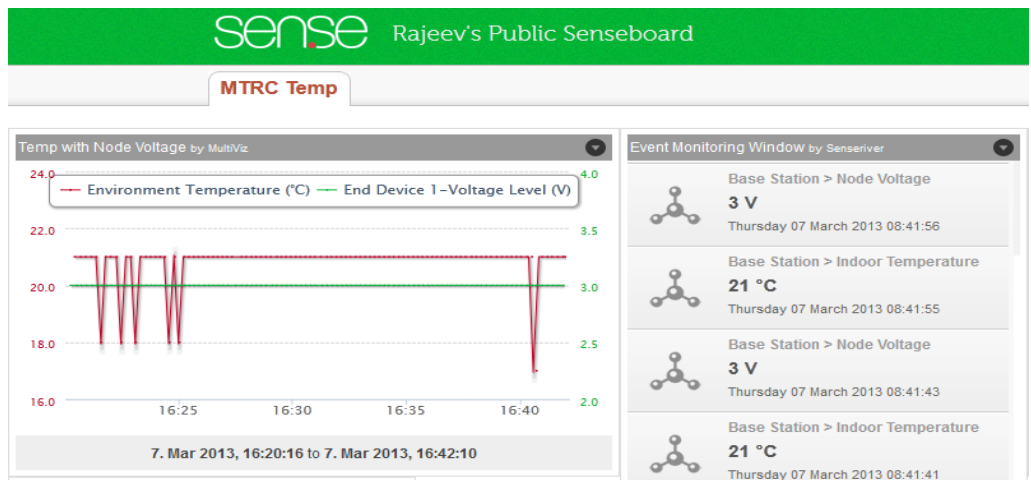

Figure 8. Senseboard displaying real-time Node Voltage and Environmental Temperature

### 6.2. Event Notification

An event notification system is also implemented on Open.Sen.se server based on measurements from sensors and predefined If-conditions. This allows monitoring End Devices supply voltage. If the voltage is too low, the End Device will enter sleep mode automatically. When Open.Sen.se server receives the voltage data for each remote Node through the base station, it compares it with a predefined threshold of 2.1V. If the measurement is equivalent to the threshold, it triggers the predefined actions. For instance, it can send a notification alert to the user via a push email or tweets. Figure 9 demonstrates the notification email received by the user as soon as the threshold value is reached. The time taken to notify the user from the time the event has occurred, in this case low battery voltage, was also measured. Using a variable DC power supply, the voltage for the End Device was manually reduced to 2.1V and the time it takes to receive the alert notification via an email was noted. Ten trials were conducted and it is observed that it takes about 8-13s and an average of 11s for the notification email to be auto generated and delivered to the user on their specified email account from the Open.Sen.se server (Figure 10). For event notification, we consider this value to be acceptable as the required time to notify the user.

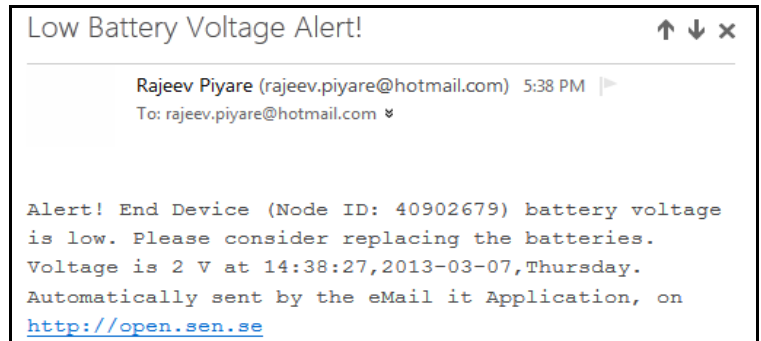

Figure 9. Notification Email to alert the user on low battery voltage

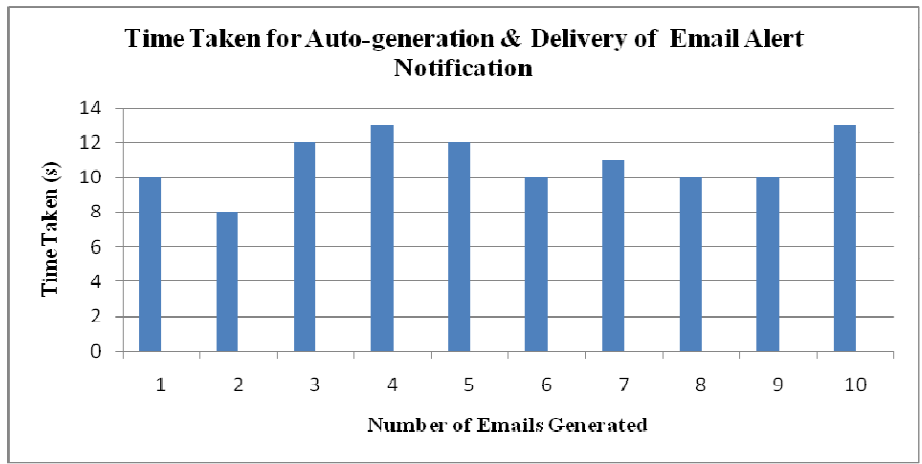

Figure 10. Time taken for auto generation and delivery of Email Alert Notification (10 attempts)

### 6.3. Battery Lifetime of the End Devices

For wireless sensor networks, energy efficiency is one of the important functional indexes since it directly affects the life cycle of the system. Replacing batteries regularly for failed sensor nodes in huge wireless networks is not convenient due to terrain and space limitations and also due to hazardous environments in which they are placed in. Therefore, the best method to save energy is setting sleep mechanism. The power consumption measurement is only carried out for the End Devices as the Coordinator is mains powered at the base station. To provide for an energy-efficient operation mode, End Devices are configured to be in a cyclic sleep mode (SM = 4). After transmission has completed, the End Device will return to sleep mode for another sleep cycle. The following Table 1 shows the average power consumption during different modes of an End Device. The measured average power consumption is not considering the power consumed by the XBee module only, but also includes the voltage regulation component and its peripheral circuits.

International Journal of Computer Networks & Communications (IJCNC) Vol.5, No.5, September 2013

Table 1.  Current measurement of an End Device

| Parameters | End Device |
|---|---|
| Activate and Deactivate current ($I_{onoff}$) | 8.1mA |
| Listen current ($I_{listen}$) | 40mA |
| Transmitter current ($I_{trans}$) | 38mA |
| Sleep current ($I_{sleep}$) | 0.6mA |
| Battery Capacity | 2000mAh |
| Battery Voltage | 9V |

The transmitted data from the End Device consisted of 2 bytes (one to encode the sensed temperature and the other for the supply voltage). With this data, MATLAB® simulations were conducted to estimate the lifetime of XBee ZB wireless sensor nodes with variable data packet size and different values of consecutive transmission time (update period) as shown in Figure 11.

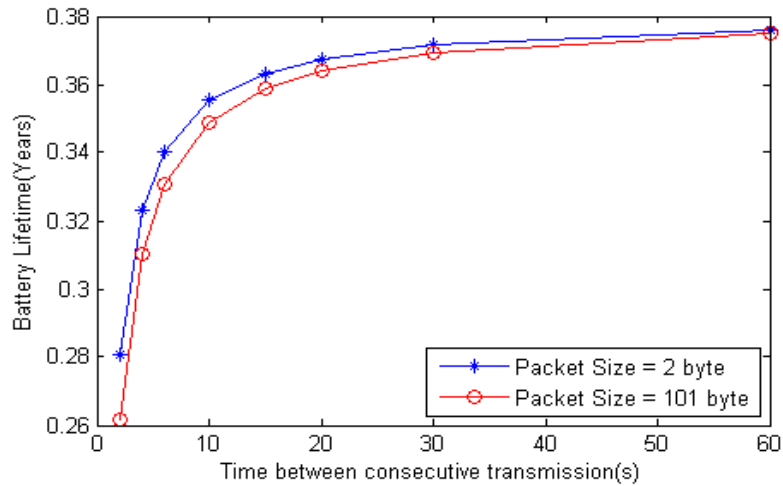

Figure 11.  Wireless Sensor Network Node lifetime with different packet size and update period

The figures include two extreme cases for the value of data size: 2 bytes and 102 bytes which is the maximum admissible value of the ZigBee/802.15.4 MAC payload. From the results obtained, the figure shows that ZigBee technology provides a typical maximum battery lifetime of up to several years for many typical scenarios of mote networks. It was also observed that the lifetime of the node decreases as the packet size increases. Hence, it is also possible to achieve longer lifetime for battery powered sensor nodes using high current capacity lithium batteries. Apparently, the power consumption of ZigBee End Devices using the cyclic sleep mode can be reduced effectively, which will improve the lifetime of the entire network.

## 7. CONCLUSIONS

This paper proposed a flexible architecture for integration of Wireless Sensor Networks to the Cloud for sensor data collection and sharing using REST based Web services as an interoperable application layer which can be directly integrated into other applications. To avoid loss of data and network disruption due to failure of Coordinator, we embedded intelligence at different architectural layers to accommodate for the diverse requirements of possible application scenarios with minimum redesign and recoding. The evaluation results illustrate that the sensor data can be accessed by the users anywhere and on any mobile device with internet access. The results also demonstrated that it takes an average of 11s for the alert notification email to be auto generated and delivered to the user on their specified email account from the Open.Sen.se server. In addition, using the sleep mechanism for low power XBee ZB transceiver modules provided an energy efficient approach to increase the lifetime of sensor nodes.

Our future research will focus on integrating Body Sensor Networks (BSNs) to the Cloud for real-time patient monitoring and notification.

## ACKNOWLEDGEMENTS

This work was supported by Priority Research Centers program through the National Research Foundation of Korea (NRF) funded by the Ministry of Education, Science and Technology (2009-0093828) and MKEC (The Ministry of Knowledge Economy), Korea, under the ITRC (Information Technology Research Center) supported program supervised by the NIPA, National IT Industry Promotion Agency (NIPA-2013-H0301-13-2005).